\documentclass[
reprint,
groupedaddress,
twocolumn,
 amsmath,amssymb,
 aps,
pra,
floatfix
]{revtex4-2}

\usepackage{textcomp}
\usepackage{graphicx}% Include figure files
\usepackage{dcolumn}% Align table columns on decimal point
\usepackage{bm}% bold math
\usepackage{float}
\usepackage[colorlinks, allcolors=blue]{hyperref}% add hypertext capabilities
\usepackage{todonotes}

\newcommand*{\figref}[2][]{\ref{#2}\hyperref[{#2}]{#1}}
\graphicspath{{./figs/}}

\begin{document}

%\preprint{APS/123-QED}

\title{Casimir-force spectroscopy of broadband optical response}

\author{C. F. Shelden}
%\email{cashelden@ucdavis.edu}
 \affiliation{Department of Electrical and Computer Engineering, University of California, Davis, CA 95616, USA}

\author{J. N. Munday}
\email{jnmunday@ucdavis.edu}
\homepage{https://mundaylab.com}
\affiliation{Department of Electrical and Computer Engineering, University of California, Davis, CA 95616, USA}

\date{\today}% It is always \today, today,

\begin{abstract}

Broadband optical response governs light–matter interactions across photonics, plasmonics, thermal radiation, and quantum fluctuation electrodynamics, yet determining a continuous dielectric function over many decades in frequency typically requires combining multiple spectroscopies, extrapolations, and material models. Here we show that quantum-fluctuation forces provide a route to broadband optical characterization. Casimir interactions depend on the dielectric response of materials across the electromagnetic spectrum, but this information is encoded through Lifshitz theory in a spectrally weighted and nontrivial way. By training physics-constrained supervised learning models on synthetic dielectric spectra and their corresponding Casimir force curves, we invert this relationship and constrain the complex permittivity of materials over more than seven orders of magnitude in frequency and provide an estimate of their values from force–distance data. The reconstruction reveals a direct separation–frequency correspondence: large separations constrain the free-carrier response, whereas shorter separations encode higher-frequency resonant structure. Applying the method to measured force gradients identifies the current experimental limits imposed by measurement noise, restricted separation range, and model complexity. These results establish fluctuation-induced forces as a spectrally weighted route to broadband optical characterization and define the experimental and physical limits that govern what spectral information is accessible from near-field quantum electromagnetic measurements.

\end{abstract}

\maketitle

%\tableofcontents

\section{Introduction}

The optical response of a material over broad frequency ranges governs light–matter interactions across photonics and plasmonics, thermal radiation, near-field energy transfer, and quantum fluctuation electrodynamics~\cite{maier_plasmonics_2007, francoeur_role_2008, khandekar_thermal_2019, lifshitz_theory_1956, dzyaloshinskii_general_1961, pirozhenko_sample_2006, svetovoy_optical_2008, van_zwol_influence_2009}. In practice, however, obtaining a continuous complex dielectric function across the far-infrared, infrared, visible, ultraviolet, and higher-frequency regimes remains challenging. Conventional optical characterization methods provide high sensitivity within particular spectral windows, but constructing broadband dielectric functions often requires combining measurements from multiple instruments, applying extrapolations outside the measured range, or imposing Kramers–Kronig-consistent model assumptions~\cite{kuzmenko_kramerskronig_2005}. These challenges are especially consequential in Casimir physics, where the measured interaction depends collectively on material response across a wide range of frequencies~\cite{pirozhenko_sample_2006, svetovoy_optical_2008, van_zwol_influence_2009, decca_measurement_2003, iannuzzi_effect_2004, klimchitskaya_casimir_2009, yang_optical_2015}.

Casimir interactions provide a fundamentally different route to probing optical response. These forces arise from quantum and thermal fluctuations of the electromagnetic field and depend on how materials modify those fluctuations through their frequency-dependent dielectric functions and electromagnetic boundary conditions~\cite{lifshitz_theory_1956, dzyaloshinskii_general_1961, casimir_attraction_1948, parsegian_van_2005, rodriguez_casimir_2011}. Unlike resonant optical probes, a Casimir force measurement does not interrogate a single narrow band of frequencies. Instead, it integrates contributions from electromagnetic fluctuations across the spectrum, with the relative weighting determined by material properties, geometry, temperature, and separation distance~\cite{lifshitz_theory_1956, dzyaloshinskii_general_1961, decca_measurement_2003, iannuzzi_effect_2004, klimchitskaya_casimir_2009}. This broadband sensitivity has enabled a range of material-dependent fluctuation phenomena, including tunable attraction and repulsion~\cite{munday_measured_2009, chen_nonequilibrium_2016, zhao_stable_2019, shelden_opportunities_2024}, anisotropy-induced torques~\cite{somers_measurement_2018, thiyam_distance-dependent_2018, antezza_giant_2020, spreng_recent_2022}, non-contact momentum transfer~\cite{manjavacas_lateral_2017, sanders_nanoscale_2019, gao_thermal_2021}, and self-assembly~\cite{munkhbat_tunable_2021, kucukoz_quantum_2024, wang_nanoalignment_2024, hoskova_casimir_2025}. 

The theoretical description of these interactions is most naturally expressed within Lifshitz theory, in which material properties enter through the dielectric function evaluated at imaginary frequencies~\cite{lifshitz_theory_1956, dzyaloshinskii_general_1961}. This formulation follows from a contour transformation that simplifies the treatment of quantum fluctuations, but it also obscures the connection between measured real-frequency optical properties and the resulting force~\cite{wick_properties_1954, rodriguez_virtual_2007}. Although the real-frequency spectrum of the Casimir effect has been discussed previously~\cite{ford_spectrum_1993, ellingsen_frequency_2008}, this perspective has not yielded a practical route for reconstructing a material’s broadband dielectric response from force measurements. In most applications, Casimir calculations instead rely on tabulated optical data, concatenated measurements, low-frequency extrapolations, or simplified oscillator models, each of which can introduce uncertainty into predicted forces~\cite{pirozhenko_sample_2006, svetovoy_optical_2008, van_zwol_influence_2009, yang_optical_2015}.

This situation raises a central question for optical characterization: can the broadband dielectric response of a material be reconstructed directly from a measurement of fluctuation-induced forces? Such an inversion is not analytically tractable because the measured force is a nonlinear, spectrally weighted functional of the dielectric response. However, the mapping is strongly constrained by electrodynamics. Here we exploit this structure by treating the inversion of Lifshitz theory as a supervised learning problem. We generate physically motivated dielectric spectra, compute their corresponding Casimir interactions, and train machine-learning models to recover the real-frequency complex permittivity from force–distance data~\cite{formisano_machine_2024, aarts_physics-driven_2025, peng_information-distilled_2025}.

We demonstrate that Casimir force curves contain sufficient information to constrain the dielectric response and provide an estimate of their values over more than seven orders of magnitude in frequency. The reconstruction reveals a clear physical correspondence between measurement separation and spectral sensitivity: large separations constrain the free-carrier response, whereas shorter separations encode higher-frequency response. We first establish the inversion for Drude-like conductive materials, then extend it to more complex Drude–Lorentz spectra and realistic optical data for metals. Finally, we apply the framework to experimentally measured Casimir force gradients and use the results to quantify the practical limits imposed by measurement noise, restricted separation range, and model complexity. Together, these results establish fluctuation-induced forces as a spectrally weighted probe of broadband optical response and define the conditions under which Casimir-force-based spectroscopy can provide useful optical information.

\section{Results and Discussion}

To determine whether fluctuation-induced forces can serve as a broadband probe of optical response, we formulate the inversion of Lifshitz theory as a supervised learning problem~\cite{formisano_machine_2024, aarts_physics-driven_2025, peng_information-distilled_2025, raissi_physics-informed_2019}. We consider the Casimir interaction between a gold sensing surface and a planar sample with unknown optical properties, using the force as a function of separation as the experimental input and the real-frequency complex permittivity of the sample as the target output. Training data are generated by constructing physically motivated dielectric spectra and computing their corresponding Casimir forces using Lifshitz theory~\cite{lifshitz_theory_1956, dzyaloshinskii_general_1961, spreng_sprengjamincaliforcia_2025}, thereby embedding the electrodynamic constraints of fluctuation-induced interactions directly into the learning process. We evaluated several supervised learning architectures and use random-forest ensemble models for the results presented here because they provided robust reconstructions across the training sets considered; model selection, parameter sampling, and hyperparameter optimization are described in the Methods and Supplementary Information. This framework allows us to determine both what optical information is encoded in force–distance data and what physical or experimental factors limit its recovery.

\begin{figure*}
	\centering
	\includegraphics[width=5.6in]{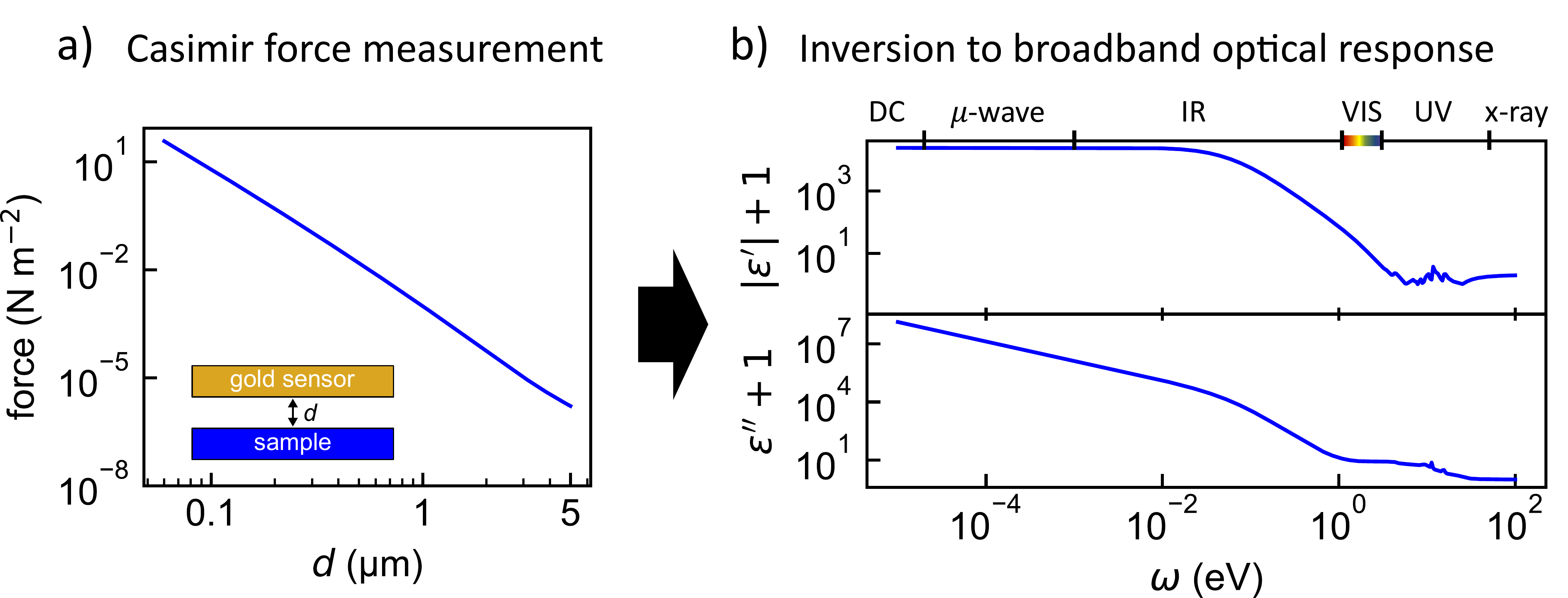}
	\caption{\textbf{Concept of fluctuation-enabled broadband optical characterization}. \textbf{a}, Representative Casimir force as a function of separation between a gold sensing surface and a planar sample with unknown optical properties. Within Lifshitz theory, the force depends on the material’s dielectric response across the electromagnetic spectrum. \textbf{b}, Schematic illustration of the inversion approach, in which supervised learning models map force–distance curves onto the real and imaginary components of the sample’s permittivity over a wide frequency range. This framework exploits the structured dependence of Casimir interactions on broadband quantum electromagnetic fluctuations to recover real-frequency optical properties from force measurements.}
	\label{fig:1}
\end{figure*}

Figure~\ref{fig:1} illustrates the concept of fluctuation-enabled optical characterization. A measured Casimir force curve between a gold sensing surface and a material of unknown optical response provides a near-field electromagnetic observable that depends on the sample’s dielectric function across a broad spectral range. Within Lifshitz theory, this dependence is not a direct sampling of the real-frequency permittivity, but a spectrally weighted functional of the response evaluated along the imaginary-frequency axis. Our approach learns the inverse relationship implied by this theory, mapping force–distance curves onto the real and imaginary components of the real-frequency permittivity. In this way, the force measurement is treated not only as a probe of the Casimir interaction, but as an indirect measurement of broadband optical response.

As a first test of the inversion framework, we consider materials whose dielectric response is described solely by a Drude model~\cite{ashcroft_solid_1976}. This simplified case captures the low-frequency electrodynamic behavior of conductive materials using a minimal set of physically interpretable parameters and provides a controlled setting in which to evaluate whether force–distance data contain sufficient information to recover the optical response. Synthetic permittivity spectra spanning a wide range of plasma frequencies are used to generate corresponding Casimir force curves, and a subset of these data is withheld for validation. Across this parameter space, the reconstructed real and imaginary components of the permittivity closely reproduce the ground truth, demonstrating that the inversion accurately captures the dominant free-carrier response encoded in the force measurements. The quality of the reconstruction is largely independent of plasma frequency, indicating that the approach is robust across materials with widely varying characteristic electronic energy scales.

\begin{figure*}
	\centering
	\includegraphics[width=5.5in]{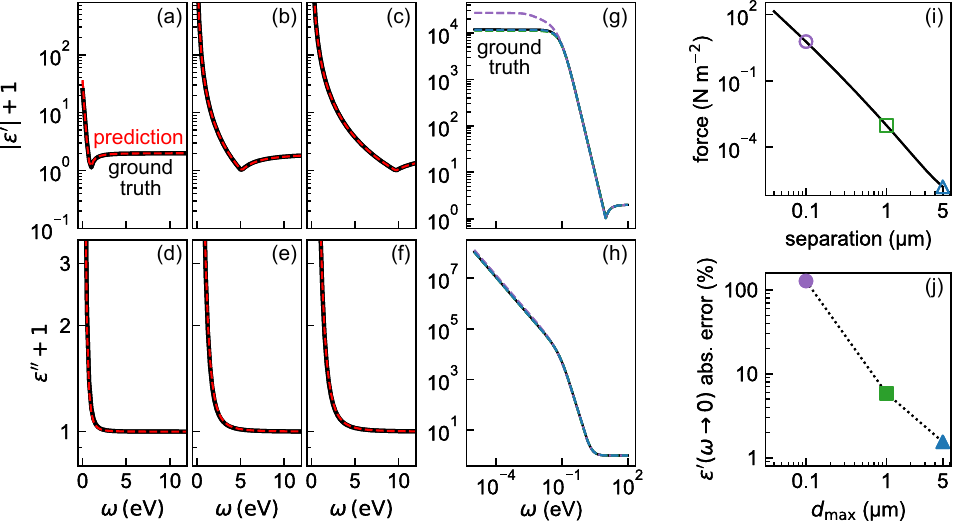}
	\caption{\textbf{Separation-dependent spectral sensitivity in Casimir-force reconstruction}. \textbf{a–f}, Reconstructed real \textbf{(a–c)} and imaginary \textbf{(d–f)} components of the permittivity for synthetic materials with widely varying plasma frequencies, demonstrating recovery of the dominant broadband optical response from Casimir force data. \textbf{g,h,} Effect of the maximum separation $d_\text{max}$ included in the force–distance data on the reconstructed real \textbf{(g)} and imaginary \textbf{(h)} permittivity. Reducing $d_\text{max}$ leads to increased deviations at low frequencies, while higher-frequency response remains comparatively unaffected. \textbf{i,} Representative Casimir force curve indicating the separation ranges used for reconstruction. \textbf{j,} Absolute error in the low-frequency limit of the real permittivity as a function of $d_\text{max}$, showing systematic improvement with increasing maximum separation. These results reveal a direct physical correspondence between measurement separation and spectral sensitivity, reflecting how long-wavelength quantum fluctuations dominate Casimir interactions at large separations.}
	\label{fig:2}
\end{figure*}

Figure~\ref{fig:2} establishes how the separation range of a force measurement determines the spectral information that can be recovered. For Drude-model materials with widely varying plasma frequencies, the reconstructed real and imaginary permittivity closely reproduce the ground truth across the frequency range considered, demonstrating that force–distance data can encode the dominant conductive optical response. The reconstruction becomes more revealing when the maximum separation is varied. Removing large-separation data primarily degrades the low-frequency response, while leaving higher-frequency features comparatively less affected. This behavior reflects the spectral weighting inherent to Lifshitz theory: long-wavelength, low-frequency fluctuations contribute most strongly at larger separations, whereas shorter separations increase sensitivity to higher-frequency response. The systematic reduction in low-frequency error with increasing maximum separation therefore provides a quantitative separation–frequency correspondence for fluctuation-based optical characterization.

Together, these results show that the inversion is not merely a numerical mapping between force curves and dielectric spectra. Instead, it reflects a physically structured encoding of optical information by quantum electromagnetic fluctuations. The accessible spectral information is controlled by the measurement geometry, particularly the separation range, providing a direct link between experimental design and the portion of the optical response that can be reconstructed.

\begin{figure*}
	\centering
	\includegraphics[width=5.5in]{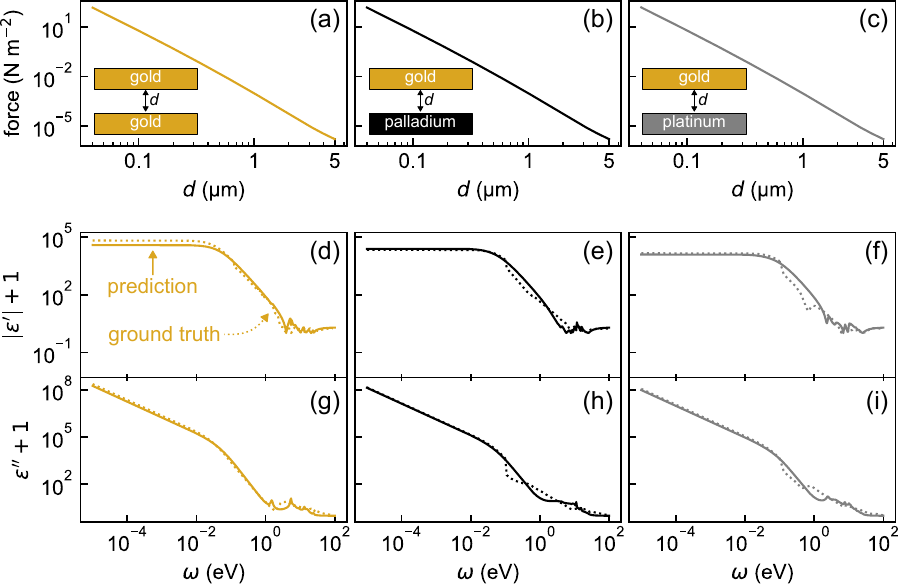}
	\caption{\textbf{Recovery of realistic metallic optical response from simulated force data.} \textbf{a–c,} Simulated Casimir force curves for gold, palladium, and platinum, generated using experimentally tabulated optical data. \textbf{d–f,} Reconstructed real component of the permittivity for each metal, compared with the corresponding ground truth. \textbf{g–i,} Reconstructed imaginary component of the permittivity. At low frequencies, where free-carrier response dominates, the reconstructed spectra capture the dominant optical response for all materials. At higher frequencies, resonant features are recovered more approximately, reflecting their weaker contribution to the Casimir interaction and the finite physical information encoded in the force. These results illustrate both the robustness and the intrinsic spectral weighting of Casimir-force-based optical reconstruction.}
	\label{fig:3}
\end{figure*}

Figure~\ref{fig:3} tests whether the inversion can recover more realistic optical response beyond ideal Drude behavior. We train on spectra containing both Drude and Lorentz contributions and apply the model to simulated Casimir force curves generated from tabulated optical data for gold, palladium, and platinum~\cite{palik_handbook_1991, rakic_optical_1998}. These materials provide a stringent test because their dielectric functions include both low-frequency free-carrier response and higher-frequency interband structure. The Drude and Drude–Lorentz training sets, including parameter ranges and frequency sampling, are described in the Methods and Supplementary Information.

The reconstructed spectra reproduce the dominant low-frequency response for all three metals, indicating that fluctuation-induced forces strongly constrain the electrodynamic response that most directly influences the Casimir interaction. Higher-frequency resonant features are recovered more approximately: their presence and approximate spectral locations are captured, but fine structure and amplitudes are not uniquely reconstructed. This behavior may be expected for a spectrally weighted inverse measurement. Features that weakly perturb the force, or that lie outside the representational capacity of the training spectra, cannot be recovered with the same fidelity as the dominant low-frequency response. We also note that the logarithmic plotting scale compresses relative differences across a wide dynamic range; deviations that appear modest on the log scale may be more apparent on a linear scale.

These results clarify the meaning of “broadband” in this context. Casimir forces are sensitive to material response over a broad frequency range, but they do not weight all frequencies equally. The inversion therefore provides a broadband, physically weighted reconstruction rather than a uniform-resolution spectrum. This distinction is important for positioning fluctuation-induced forces as complementary to conventional optical spectroscopy. Conventional optical methods can provide high precision within specific spectral windows, whereas force-based reconstruction probes the integrated optical response that directly governs fluctuation electrodynamic interactions.

\begin{figure*}
	\centering
	\includegraphics[width=4.51in]{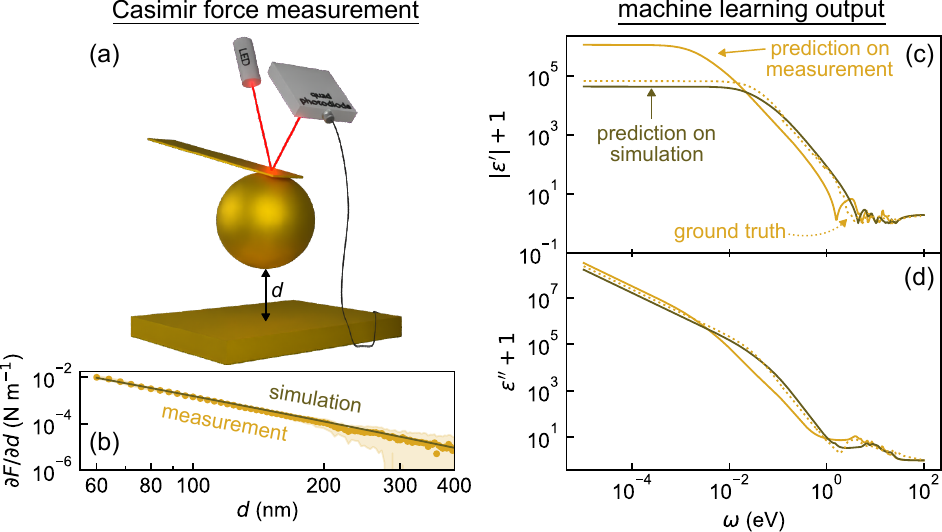}
	\caption{\textbf{Application of the inversion framework to experimentally measured Casimir force gradients.} \textbf{a,} Schematic of the atomic force microscope geometry used to measure the Casimir force gradient between a gold-coated sphere and a gold plate. \textbf{b,} Measured and simulated Casimir force gradients over the same separation range (60–400 nm). \textbf{c,d,} Reconstructed real \textbf{(c)} and imaginary \textbf{(d)} components of the permittivity of gold obtained from simulated and measured force gradients. Reconstructions based on simulated data closely reproduce the low-frequency optical response, whereas experimental data exhibit larger deviations due to noise, bias, and restricted separation access. These results delineate the experimental conditions under which Casimir-force-based optical reconstruction is viable and highlight the role of measurement precision and separation range in determining spectral sensitivity.}
	\label{fig:4}
\end{figure*}

Figure~\ref{fig:4} applies the inversion framework to experimentally measured Casimir force gradients, providing a test of fluctuation-based optical reconstruction under realistic measurement conditions. We use force-gradient data measured between a gold-coated sphere and a gold plate using an amplitude-modulated atomic-force-microscopy scheme~\cite{garrett_sensitivity_2019}, and convert the sphere–plate gradient to an equivalent plate–plate interaction using the proximity force approximation~\cite{derjaguin_theorie_1934}. To separate the effects of limited separation range from experimental noise and bias, we compare reconstructions from measured data with reconstructions from simulated force gradients over the same separation interval.

When applied to simulated force-gradient data spanning 60–400 nm, the inversion recovers the dominant low-frequency optical response, showing that this separation range retains useful spectral information. Reconstructions based on measured force gradients show larger deviations, especially in the real part of the permittivity. These deviations reflect the ill-conditioned nature of the inverse problem: small systematic errors or noise in the force curve can be amplified when mapped onto a broadband dielectric function. Differences between the optical properties of the measured sample, the probe coating, and the tabulated or independently measured reference data may also contribute.

Rather than representing a failure of the concept, this comparison identifies the experimental requirements for Casimir-force-based optical characterization. Improved force-gradient precision, broader separation ranges, better characterization of probe optical properties, and inversion models trained with realistic noise and uncertainty will be needed to move from proof-of-concept reconstruction toward quantitative spectroscopy. The experimental result therefore delineates the present operating regime of the method and highlights the measurements most important for improving spectral recovery.

\begin{figure*}
	\centering
	\includegraphics[width=5.77in]{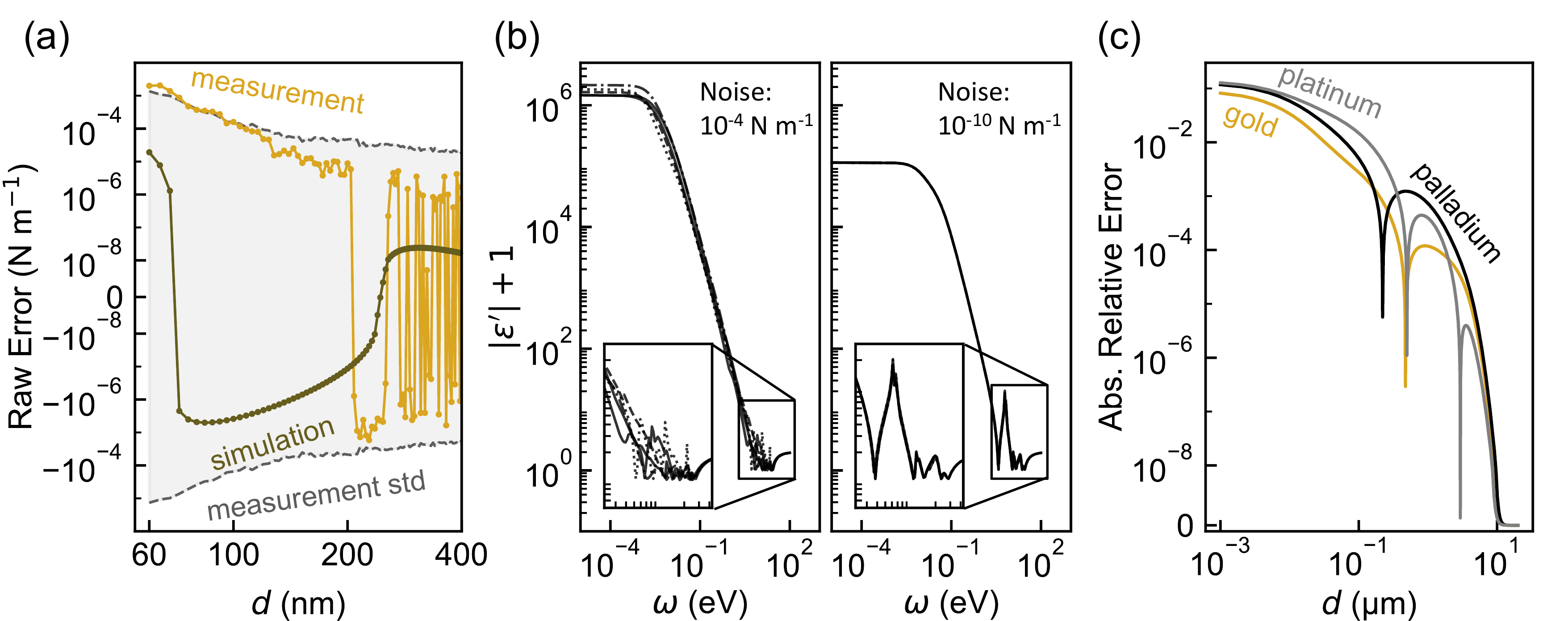}
	\caption{\textbf{Characterizing the current limits imposed by measurement uncertainty and machine learning complexity.} \textbf{a,} Error in the Casimir force gradient calculated using the predicted spectra (from measurement and simulation) shown in Fig.~\figref[d]{fig:4} relative to the original force gradients shown in Fig.~\figref[b]{fig:4}. \textbf{b,} Permittivity spectra predicted from the simulated force gradient between two gold surfaces with an added Gaussian noise amplitude of $10^{-4}\,\text{N\,m}^{-1}$ (left) and $10^{-10}\,\text{N\,m}^{-1}$ (right). A random seed is used to generate five unique noisy force gradient curves, and five permittivity spectra are generated, respectively, demonstrating that convergence of the predictions to a stable estimate is achieved for sufficiently small noise $(10^{-10}\,\text{N\,m}^{-1})$. \textbf{c,} Error in the Casimir force calculated using the predicted spectra shown in Fig.~\figref[g-i]{fig:2} relative to the original forces shown in Fig.~\figref[a-c]{fig:2} divided by the original forces.}
	\label{fig:5}
\end{figure*}

Figure~\ref{fig:5} quantifies how measurement uncertainty and model complexity limit the reconstructed dielectric response. Recalculating the Casimir force from the permittivity spectra reconstructed from measured data produces force gradients that lie within the experimental uncertainty, showing that multiple dielectric spectra can be consistent with the measured force at the current noise level. This non-uniqueness is a central feature of the inverse problem and motivates the need for uncertainty-aware reconstruction.

The effect of measurement uncertainty is further illustrated by adding controlled Gaussian noise to simulated force-gradient curves. In Fig.~\figref[b]{fig:5}, we see that the real component of the predicted permittivity varies greatly across five iterations of noisy force curve generation with a Gaussian noise amplitude of $10^{-4}\,\text{N\,m}^{-1}$ (Fig.~\figref[b]{fig:5}, left). As the noise amplitude is reduced to $10^{-10}\,\text{N\,m}^{-1}$ (~\figref[b]{fig:5}, right), the predictions converge to a constrained estimate, even for the resonant portions of the spectrum. $10^{-10}\,\text{N\,m}^{-1}$ is smaller than the standard deviation present in our measurements $(\sim$$10^{-7}\,\text{N\,m}^{-1})$; however, it is theoretically possible to push the minimum detectable force gradient towards $10^{-10}\,\text{N\,m}^{-1}$ by increasing the measurement integration time or cooling the measurement apparatus to mK temperatures~\cite{garrett_sensitivity_2019}. These results show that improved force sensitivity directly increases the optical information that can be extracted. The remaining deviations observed when reconstructing forces from predicted spectra without added measurement noise reflect the finite capacity of the current training set and model architecture, particularly for higher-frequency resonant structure.

These calculations define a path forward. Experimentally, lower-noise measurements, longer integration times, cryogenic or vacuum environments, and extended separation ranges should improve spectral recovery. Because the low-frequency response is the most robustly constrained part of the reconstruction, future uncertainty-aware implementations may also provide a way to compare competing low-frequency extrapolations, including Drude and plasma-like models. Computationally, training on broader classes of causal dielectric functions, incorporating realistic noise models, using more advanced machine learning models, and returning confidence intervals rather than single spectra would make the reconstruction more robust. Thus, the present implementation should be viewed as a first realization of fluctuation-induced optical reconstruction, with clear routes toward improved quantitative performance.

Overall, the results establish fluctuation-induced forces as a spectrally weighted probe of broadband optical response. The method does not provide uniform sensitivity across all frequencies, nor does it replace conventional optical spectroscopy in spectral regions where direct optical measurements are available with high precision. Instead, it provides a complementary measurement principle: the force directly encodes the broadband material response relevant to fluctuation electrodynamics, and the inversion framework determines which parts of that response can be recovered under realistic experimental constraints. This perspective transforms Casimir interactions from a phenomenon that depends on optical properties into a tool for interrogating them, while also defining the physical and practical limits of quantum-fluctuation-based optical characterization.

\section{Conclusion}

The broadband optical response of a material governs a wide range of photonic, plasmonic, thermal, and fluctuation-induced phenomena, yet reconstructing a continuous complex dielectric function across many decades in frequency remains challenging. Here we have shown that Casimir force measurements provide a route to probing this response through quantum electromagnetic fluctuations. By inverting Lifshitz theory using supervised learning models trained on physically motivated dielectric spectra, we reconstruct a constrained estimate of the real-frequency complex permittivity of materials over more than seven orders of magnitude in frequency from force–distance data.

The reconstruction reveals that fluctuation-induced forces encode optical response in a structured and physically interpretable way. Force measurements at larger separations preferentially constrain the aggregate free-carrier response, whereas shorter separations encode higher-frequency contributions. This separation–frequency correspondence establishes Casimir interactions as a spectrally weighted probe of broadband dielectric response, rather than an opaque integral over electromagnetic modes.

Extending the approach to realistic metal spectra and experimentally measured force gradients defines both the promise and the current limits of Casimir-force-based optical characterization. The dominant low-frequency response is recovered robustly, while higher-frequency resonant structure is reconstructed more approximately because it contributes more weakly to the measured force and may lie outside the representational capacity of the current training set. Experimental noise, restricted separation range, uncertainty in probe and sample optical properties, and model complexity further limit the information that can be extracted from present measurements.

These limitations point to clear routes for improvement. Lower-noise force measurements, broader separation ranges, better characterization of probe materials, training sets based on broader classes of causal dielectric functions, improved machine learning models, and uncertainty-aware inversion methods could all improve the fidelity and reliability of the reconstructed spectra. More broadly, the framework developed here provides a way to determine not only whether optical information is present in fluctuation-induced forces, but also which spectral features are recoverable under realistic experimental constraints.

Together, these results establish fluctuation-induced forces as a complementary route to broadband optical characterization. Rather than replacing conventional optical spectroscopy, Casimir-force-based reconstruction probes the broadband material response that directly governs fluctuation electrodynamic interactions. This perspective transforms Casimir forces from a phenomenon that depends on optical properties into a tool for interrogating them, opening opportunities for quantum-fluctuation-based measurements of optical response in materials, geometries, and spectral regimes that are difficult to access by conventional means.

\section*{Acknowledgments}

The authors acknowledge financial support from the Defense Advanced Research Projects Agency (DARPA) QUEST Projects Contract No. HR00112090084. C.F.S. acknowledges support from the National Science Foundation Graduate Research Fellowship Program under Grant No. 2036201. Any opinions, findings, and conclusions or recommendations expressed in this material are those of the author(s) and do not necessarily reflect the views of the National Science Foundation or DARPA.

\bibliography{references}

@article{aarts_physics-driven_2025,
	title = {Physics-driven learning for inverse problems in quantum chromodynamics},
	copyright = {2025 Springer Nature Limited},
	issn = {2522-5820},
	url = {https://www.nature.com/articles/s42254-024-00798-x},
	doi = {10.1038/s42254-024-00798-x},
	urldate = {2025-01-07},
	journal = {Nature Reviews Physics},
	publisher = {Nature Publishing Group},
	author = {Aarts, Gert and Fukushima, Kenji and Hatsuda, Tetsuo and Ipp, Andreas and Shi, Shuzhe and Wang, Lingxiao and Zhou, Kai},
	month = jan,
	year = {2025},
	pages = {1--10},
}

@article{shelden_opportunities_2024,
	title = {Opportunities and challenges involving repulsive {Casimir} forces in nanotechnology},
	volume = {11},
	copyright = {All rights reserved},
	issn = {1931-9401},
	url = {https://doi.org/10.1063/5.0218274},
	doi = {10.1063/5.0218274},
	number = {4},
	urldate = {2024-12-03},
	journal = {Applied Physics Reviews},
	author = {Shelden, C. and Spreng, B. and Munday, J. N.},
	month = dec,
	year = {2024},
	pages = {041325},
}

@article{dzyaloshinskii_general_1961,
	title = {The {General} {Theory} of {Van} der {Waals} {Forces}},
	volume = {10},
	issn = {0001-8732},
	url = {https://doi.org/10.1080/00018736100101281},
	doi = {10.1080/00018736100101281},
	number = {38},
	urldate = {2022-08-02},
	journal = {Advances in Physics},
	publisher = {Taylor \& Francis},
	author = {Dzyaloshinskii, I. E. and Lifshitz, E. M. and Pitaevskii, L.P.},
	month = apr,
	year = {1961},
	note = {\_eprint: https://doi.org/10.1080/00018736100101281},
	pages = {165--209},
}

@article{chen_nonequilibrium_2016,
	title = {Nonequilibrium {Casimir} {Force} with a {Nonzero} {Chemical} {Potential} for {Photons}},
	volume = {117},
	issn = {0031-9007, 1079-7114},
	url = {https://link.aps.org/doi/10.1103/PhysRevLett.117.267401},
	doi = {10.1103/PhysRevLett.117.267401},
	number = {26},
	urldate = {2021-05-08},
	journal = {Physical Review Letters},
	author = {Chen, Kaifeng and Fan, Shanhui},
	month = dec,
	year = {2016},
	pages = {267401},
}

@article{munday_measured_2009,
	title = {Measured long-range repulsive {Casimir}–{Lifshitz} forces},
	volume = {457},
	issn = {0028-0836, 1476-4687},
	url = {http://www.nature.com/articles/nature07610},
	doi = {10.1038/nature07610},
	number = {7226},
	urldate = {2021-05-08},
	journal = {Nature},
	author = {Munday, J. N. and Capasso, Federico and Parsegian, V. Adrian},
	month = jan,
	year = {2009},
	pages = {170--173},
}

@article{francoeur_role_2008,
	title = {Role of fluctuational electrodynamics in near-field radiative heat transfer},
	volume = {109},
	issn = {00224073},
	url = {https://linkinghub.elsevier.com/retrieve/pii/S0022407307002427},
	doi = {10.1016/j.jqsrt.2007.08.017},
	number = {2},
	urldate = {2021-05-08},
	journal = {Journal of Quantitative Spectroscopy and Radiative Transfer},
	author = {Francoeur, Mathieu and Pinar Mengüç, M.},
	month = jan,
	year = {2008},
	pages = {280--293},
}

@article{zhao_stable_2019,
	title = {Stable {Casimir} equilibria and quantum trapping},
	volume = {364},
	issn = {0036-8075, 1095-9203},
	url = {https://www.sciencemag.org/lookup/doi/10.1126/science.aax0916},
	doi = {10.1126/science.aax0916},
	number = {6444},
	urldate = {2021-05-05},
	journal = {Science},
	author = {Zhao, Rongkuo and Li, Lin and Yang, Sui and Bao, Wei and Xia, Yang and Ashby, Paul and Wang, Yuan and Zhang, Xiang},
	month = jun,
	year = {2019},
	pages = {984--987},
}

@article{somers_measurement_2018,
	title = {Measurement of the {Casimir} torque},
	volume = {564},
	issn = {0028-0836, 1476-4687},
	url = {http://www.nature.com/articles/s41586-018-0777-8},
	doi = {10.1038/s41586-018-0777-8},
	number = {7736},
	urldate = {2021-05-08},
	journal = {Nature},
	author = {Somers, David A. T. and Garrett, Joseph L. and Palm, Kevin J. and Munday, Jeremy N.},
	month = dec,
	year = {2018},
	pages = {386--389},
}

@article{garrett_sensitivity_2019,
	title = {Sensitivity and accuracy of {Casimir} force measurements in air},
	volume = {100},
	issn = {2469-9926, 2469-9934},
	url = {https://link.aps.org/doi/10.1103/PhysRevA.100.022508},
	doi = {10.1103/PhysRevA.100.022508},
	number = {2},
	urldate = {2021-05-05},
	journal = {Physical Review A},
	author = {Garrett, Joseph L. and Somers, David A. T. and Sendgikoski, Kyle and Munday, Jeremy N.},
	month = aug,
	year = {2019},
	pages = {022508},
}

@book{parsegian_van_2005,
	edition = {1},
	title = {Van der {Waals} {Forces}: {A} {Handbook} for {Biologists}, {Chemists}, {Engineers}, and {Physicists}},
	isbn = {978-0-521-54778-9 978-0-521-83906-8 978-0-511-61460-6},
	shorttitle = {Van der {Waals} {Forces}},
	url = {https://www.cambridge.org/core/product/identifier/9780511614606/type/book},
	doi = {10.1017/CBO9780511614606},
	urldate = {2021-05-05},
	publisher = {Cambridge University Press},
	author = {Parsegian, V. Adrian},
	month = nov,
	year = {2005},
}

@article{thiyam_distance-dependent_2018,
	title = {Distance-{Dependent} {Sign} {Reversal} in the {Casimir}-{Lifshitz} {Torque}},
	volume = {120},
	url = {https://link.aps.org/doi/10.1103/PhysRevLett.120.131601},
	doi = {10.1103/PhysRevLett.120.131601},
	number = {13},
	urldate = {2021-08-21},
	journal = {Physical Review Letters},
	publisher = {American Physical Society},
	author = {Thiyam, Priyadarshini and Parashar, Prachi and Shajesh, K. V. and Malyi, Oleksandr I. and Boström, Mathias and Milton, Kimball A. and Brevik, Iver and Persson, Clas},
	month = mar,
	year = {2018},
	pages = {131601},
}

@article{iannuzzi_effect_2004,
	chapter = {Physical Sciences},
	title = {Effect of hydrogen-switchable mirrors on the {Casimir} force},
	volume = {101},
	copyright = {Copyright © 2004, The National Academy of Sciences},
	issn = {0027-8424, 1091-6490},
	url = {https://www.pnas.org/content/101/12/4019},
	doi = {10.1073/pnas.0400876101},
	number = {12},
	urldate = {2021-08-19},
	journal = {Proceedings of the National Academy of Sciences},
	publisher = {National Academy of Sciences},
	author = {Iannuzzi, Davide and Lisanti, Mariangela and Capasso, Federico},
	month = mar,
	year = {2004},
	pages = {4019--4023},
}

@article{decca_measurement_2003,
	title = {Measurement of the {Casimir} {Force} between {Dissimilar} {Metals}},
	volume = {91},
	url = {https://link.aps.org/doi/10.1103/PhysRevLett.91.050402},
	doi = {10.1103/PhysRevLett.91.050402},
	number = {5},
	urldate = {2023-04-29},
	journal = {Physical Review Letters},
	publisher = {American Physical Society},
	author = {Decca, R. S. and López, D. and Fischbach, E. and Krause, D. E.},
	month = jul,
	year = {2003},
	pages = {050402},
}

@article{rodriguez_virtual_2007,
	title = {Virtual photons in imaginary time: {Computing} exact {Casimir} forces via standard numerical electromagnetism techniques},
	volume = {76},
	shorttitle = {Virtual photons in imaginary time},
	url = {https://link.aps.org/doi/10.1103/PhysRevA.76.032106},
	doi = {10.1103/PhysRevA.76.032106},
	number = {3},
	urldate = {2023-02-25},
	journal = {Physical Review A},
	publisher = {American Physical Society},
	author = {Rodriguez, Alejandro and Ibanescu, Mihai and Iannuzzi, Davide and Joannopoulos, J. D. and Johnson, Steven G.},
	month = sep,
	year = {2007},
	pages = {032106},
}

@article{derjaguin_theorie_1934,
	title = {Theorie des {Anhaftens} kleiner {Teilchen}},
	volume = {69},
	issn = {1435-1536},
	url = {https://doi.org/10.1007/BF01433225},
	doi = {10.1007/BF01433225},
	number = {2},
	urldate = {2023-02-20},
	journal = {Kolloid-Zeitschrift},
	author = {Derjaguin, B.},
	month = nov,
	year = {1934},
	pages = {155--164},
}

@article{casimir_attraction_1948,
	title = {On the {Attraction} {Between} {Two} {Perfectly} {Conducting} {Plates}},
	volume = {51},
	journal = {Proc. K. Ned. Akad. Wet.},
	author = {Casimir, H. B. G.},
	year = {1948},
	pages = {793--795},
}

@article{rodriguez_casimir_2011,
	title = {The {Casimir} effect in microstructured geometries},
	volume = {5},
	copyright = {2011 Nature Publishing Group, a division of Macmillan Publishers Limited. All Rights Reserved.},
	issn = {1749-4893},
	url = {https://www.nature.com/articles/nphoton.2011.39},
	doi = {10.1038/nphoton.2011.39},
	number = {4},
	urldate = {2022-08-08},
	journal = {Nature Photonics},
	publisher = {Nature Publishing Group},
	author = {Rodriguez, Alejandro W. and Capasso, Federico and Johnson, Steven G.},
	month = apr,
	year = {2011},
	note = {Number: 4},
	pages = {211--221},
}

@article{munkhbat_tunable_2021,
	title = {Tunable self-assembled {Casimir} microcavities and polaritons},
	volume = {597},
	copyright = {2021 The Author(s), under exclusive licence to Springer Nature Limited},
	issn = {1476-4687},
	url = {https://www.nature.com/articles/s41586-021-03826-3},
	doi = {10.1038/s41586-021-03826-3},
	number = {7875},
	urldate = {2022-08-07},
	journal = {Nature},
	publisher = {Nature Publishing Group},
	author = {Munkhbat, Battulga and Canales, Adriana and Küçüköz, Betül and Baranov, Denis G. and Shegai, Timur O.},
	month = sep,
	year = {2021},
	note = {Number: 7875},
	pages = {214--219},
}

@book{ashcroft_solid_1976,
	title = {Solid {State} {Physics}},
	publisher = {Holt-Saunders},
	author = {Ashcroft, Neil W. and Mermin, David N.},
	year = {1976},
}

@article{spreng_recent_2022,
	title = {Recent developments on the {Casimir} torque},
	volume = {37},
	issn = {0217-751X},
	url = {https://www.worldscientific.com/doi/10.1142/S0217751X22410111},
	doi = {10.1142/S0217751X22410111},
	number = {19},
	urldate = {2024-01-31},
	journal = {International Journal of Modern Physics A},
	publisher = {World Scientific Publishing Co.},
	author = {Spreng, Benjamin and Gong, Tao and Munday, Jeremy N.},
	month = jul,
	year = {2022},
	pages = {2241011},
}

@book{palik_handbook_1991,
	address = {San Diego, UNITED STATES},
	title = {Handbook of {Optical} {Constants} of {Solids}: {Volume} 2},
	isbn = {978-0-08-055630-7},
	shorttitle = {Handbook of {Optical} {Constants} of {Solids}},
	url = {http://ebookcentral.proquest.com/lib/ucdavis/detail.action?docID=328564},
	urldate = {2023-11-22},
	publisher = {Elsevier Science \& Technology},
	author = {Palik, Edward D.},
	year = {1991},
}

@article{lifshitz_theory_1956,
	title = {The {Theory} of {Molecular} {Attractive} {Forces} between {Solids}},
	volume = {29},
	journal = {J. Exper. Theoret. Phys. USSR},
	author = {Lifshitz, E. M.},
	year = {1956},
	pages = {94--110},
}

@article{peng_information-distilled_2025,
	title = {Information-distilled physics informed deep learning for high order differential inverse problems with extreme discontinuities},
	volume = {4},
	copyright = {2025 The Author(s)},
	issn = {2731-3395},
	url = {https://www.nature.com/articles/s44172-025-00476-5},
	doi = {10.1038/s44172-025-00476-5},
	number = {1},
	urldate = {2025-11-17},
	journal = {Communications Engineering},
	publisher = {Nature Publishing Group},
	author = {Peng, Mingsheng and Tang, Hesheng},
	month = sep,
	year = {2025},
	pages = {150},
}

@article{formisano_machine_2024,
	title = {Machine {Learning} {Approaches} for {Inverse} {Problems} and {Optimal} {Design} in {Electromagnetism}},
	volume = {13},
	copyright = {http://creativecommons.org/licenses/by/3.0/},
	issn = {2079-9292},
	url = {https://www.mdpi.com/2079-9292/13/7/1167},
	doi = {10.3390/electronics13071167},
	number = {7},
	urldate = {2025-11-17},
	journal = {Electronics},
	publisher = {Multidisciplinary Digital Publishing Institute},
	author = {Formisano, Alessandro and Tucci, Mauro},
	month = jan,
	year = {2024},
	pages = {1167},
}

@article{rakic_optical_1998,
	title = {Optical properties of metallic films for vertical-cavity optoelectronic devices},
	volume = {37},
	copyright = {© 1998 Optical Society of America},
	issn = {2155-3165},
	url = {https://opg.optica.org/ao/abstract.cfm?uri=ao-37-22-5271},
	doi = {10.1364/AO.37.005271},
	number = {22},
	urldate = {2025-10-29},
	journal = {Applied Optics},
	publisher = {Optica Publishing Group},
	author = {Rakić, Aleksandar D. and Djurišić, Aleksandra B. and Elazar, Jovan M. and Majewski, Marian L.},
	month = aug,
	year = {1998},
	pages = {5271--5283},
}

@misc{spreng_sprengjamincaliforcia_2025,
	title = {sprengjamin/califorcia},
	copyright = {MIT},
	url = {https://github.com/sprengjamin/califorcia},
	urldate = {2025-10-16},
	author = {Spreng, Benjamin},
	month = apr,
	year = {2025},
	note = {original-date: 2022-04-08T21:31:49Z},
}

@article{klimchitskaya_casimir_2009,
	title = {The {Casimir} force between real materials: {Experiment} and theory},
	volume = {81},
	shorttitle = {The {Casimir} force between real materials},
	url = {https://link.aps.org/doi/10.1103/RevModPhys.81.1827},
	doi = {10.1103/RevModPhys.81.1827},
	number = {4},
	urldate = {2024-10-02},
	journal = {Reviews of Modern Physics},
	publisher = {American Physical Society},
	author = {Klimchitskaya, G. L. and Mohideen, U. and Mostepanenko, V. M.},
	month = dec,
	year = {2009},
	pages = {1827--1885},
}

@article{kucukoz_quantum_2024,
	title = {Quantum trapping and rotational self-alignment in triangular {Casimir} microcavities},
	volume = {10},
	url = {https://www.science.org/doi/10.1126/sciadv.adn1825},
	doi = {10.1126/sciadv.adn1825},
	number = {17},
	urldate = {2024-05-06},
	journal = {Science Advances},
	publisher = {American Association for the Advancement of Science},
	author = {Küçüköz, Betül and Kotov, Oleg V. and Canales, Adriana and Polyakov, Alexander Yu. and Agrawal, Abhay V. and Antosiewicz, Tomasz J. and Shegai, Timur O.},
	month = apr,
	year = {2024},
	pages = {eadn1825},
}

@article{antezza_giant_2020,
	title = {Giant {Casimir} {Torque} between {Rotated} {Gratings} and the $\theta$=0 {Anomaly}},
	volume = {124},
	url = {https://link.aps.org/doi/10.1103/PhysRevLett.124.013903},
	doi = {10.1103/PhysRevLett.124.013903},
	number = {1},
	urldate = {2026-01-13},
	journal = {Physical Review Letters},
	publisher = {American Physical Society},
	author = {Antezza, Mauro and Chan, H. B. and Guizal, Brahim and Marachevsky, Valery N. and Messina, Riccardo and Wang, Mingkang},
	month = jan,
	year = {2020},
	pages = {013903},
}

@article{hoskova_casimir_2025,
	title = {Casimir self-assembly: {A} platform for measuring nanoscale surface interactions in liquids},
	volume = {122},
	shorttitle = {Casimir self-assembly},
	url = {https://www.pnas.org/doi/10.1073/pnas.2505144122},
	doi = {10.1073/pnas.2505144122},
	number = {31},
	urldate = {2026-01-13},
	journal = {Proceedings of the National Academy of Sciences},
	publisher = {Proceedings of the National Academy of Sciences},
	author = {Hošková, Michaela and Kotov, Oleg V. and Küçüköz, Betül and Murphy, Catherine J. and Shegai, Timur O.},
	month = aug,
	year = {2025},
	pages = {e2505144122},
}

@article{wang_nanoalignment_2024,
	title = {Nanoalignment by critical {Casimir} torques},
	volume = {15},
	copyright = {2024 The Author(s)},
	issn = {2041-1723},
	url = {https://www.nature.com/articles/s41467-024-49220-1},
	doi = {10.1038/s41467-024-49220-1},
	number = {1},
	urldate = {2026-01-13},
	journal = {Nature Communications},
	publisher = {Nature Publishing Group},
	author = {Wang, Gan and Nowakowski, Piotr and Farahmand Bafi, Nima and Midtvedt, Benjamin and Schmidt, Falko and Callegari, Agnese and Verre, Ruggero and Käll, Mikael and Dietrich, S. and Kondrat, Svyatoslav and Volpe, Giovanni},
	month = jun,
	year = {2024},
	pages = {5086},
}

@article{pirozhenko_sample_2006,
	title = {Sample dependence of the {Casimir} force},
	volume = {8},
	issn = {1367-2630},
	url = {https://doi.org/10.1088/1367-2630/8/10/238},
	doi = {10.1088/1367-2630/8/10/238},
	number = {10},
	urldate = {2026-01-13},
	journal = {New Journal of Physics},
	author = {Pirozhenko, I and Lambrecht, A and Svetovoy, V B},
	month = oct,
	year = {2006},
	pages = {238},
}

@article{svetovoy_optical_2008,
	title = {Optical properties of gold films and the {Casimir} force},
	volume = {77},
	url = {https://link.aps.org/doi/10.1103/PhysRevB.77.035439},
	doi = {10.1103/PhysRevB.77.035439},
	number = {3},
	urldate = {2026-01-13},
	journal = {Physical Review B},
	publisher = {American Physical Society},
	author = {Svetovoy, V. B. and van Zwol, P. J. and Palasantzas, G. and De Hosson, J. Th. M.},
	month = jan,
	year = {2008},
	pages = {035439},
}

@article{van_zwol_influence_2009,
	title = {Influence of dielectric properties on van der {Waals}/{Casimir} forces in solid-liquid systems},
	volume = {79},
	url = {https://link.aps.org/doi/10.1103/PhysRevB.79.195428},
	doi = {10.1103/PhysRevB.79.195428},
	number = {19},
	urldate = {2026-01-13},
	journal = {Physical Review B},
	publisher = {American Physical Society},
	author = {van Zwol, P. J. and Palasantzas, G. and De Hosson, J. Th. M.},
	month = may,
	year = {2009},
	pages = {195428},
}

@article{wick_properties_1954,
	title = {Properties of {Bethe}-{Salpeter} {Wave} {Functions}},
	volume = {96},
	doi = {10.1103/PhysRev.96.1124},
	number = {4},
	journal = {Physical Review},
	author = {Wick, G. C.},
	year = {1954},
	pages = {1124--1134},
}

@article{yang_optical_2015,
	title = {Optical dielectric function of silver},
	volume = {91},
	url = {https://link.aps.org/doi/10.1103/PhysRevB.91.235137},
	doi = {10.1103/PhysRevB.91.235137},
	number = {23},
	urldate = {2026-05-09},
	journal = {Physical Review B},
	publisher = {American Physical Society},
	author = {Yang, Honghua U. and D'Archangel, Jeffrey and Sundheimer, Michael L. and Tucker, Eric and Boreman, Glenn D. and Raschke, Markus B.},
	month = jun,
	year = {2015},
	pages = {235137},
}

@article{ford_spectrum_1993,
	title = {Spectrum of the {Casimir} effect and the {Lifshitz} theory},
	volume = {48},
	doi = {10.1103/PhysRevA.48.2962},
	number = {4},
	journal = {Physical Review A},
	author = {Ford, L. H.},
	year = {1993},
	pages = {2962--2967},
}

@article{ellingsen_frequency_2008,
	title = {Frequency spectrum of the {Casimir} force: {Interpretation} and a paradox},
	volume = {82},
	issn = {0295-5075},
	shorttitle = {Frequency spectrum of the {Casimir} force},
	url = {https://doi.org/10.1209/0295-5075/82/53001},
	doi = {10.1209/0295-5075/82/53001},
	number = {5},
	urldate = {2026-05-09},
	journal = {Europhysics Letters},
	author = {Ellingsen, S. A.},
	month = may,
	year = {2008},
	pages = {53001},
}

@book{maier_plasmonics_2007,
	address = {New York},
	title = {Plasmonics: fundamentals and applications},
	isbn = {978-0-387-37825-1},
	shorttitle = {Plasmonics},
	publisher = {Springer},
	author = {Maier, Stefan A.},
	year = {2007},
}

@article{khandekar_thermal_2019,
	title = {Thermal spin photonics in the near-field of nonreciprocal media},
	volume = {21},
	issn = {1367-2630},
	url = {https://doi.org/10.1088/1367-2630/ab494d},
	doi = {10.1088/1367-2630/ab494d},
	number = {10},
	urldate = {2026-08-29},
	journal = {New Journal of Physics},
	publisher = {IOP Publishing},
	author = {Khandekar, Chinmay and Jacob, Zubin},
	month = oct,
	year = {2019},
	pages = {103030},
}

@article{kuzmenko_kramerskronig_2005,
	title = {Kramers–{Kronig} constrained variational analysis of optical spectra},
	volume = {76},
	issn = {0034-6748},
	url = {https://doi.org/10.1063/1.1979470},
	doi = {10.1063/1.1979470},
	number = {8},
	urldate = {2026-08-29},
	journal = {Review of Scientific Instruments},
	author = {Kuzmenko, A. B.},
	month = jul,
	year = {2005},
	pages = {083108},
}

@article{manjavacas_lateral_2017,
	title = {Lateral {Casimir} {Force} on a {Rotating} {Particle} near a {Planar} {Surface}},
	volume = {118},
	url = {https://link.aps.org/doi/10.1103/PhysRevLett.118.133605},
	doi = {10.1103/PhysRevLett.118.133605},
	number = {13},
	urldate = {2026-08-29},
	journal = {Physical Review Letters},
	publisher = {American Physical Society},
	author = {Manjavacas, Alejandro and Rodríguez-Fortuño, Francisco J. and García de Abajo, F. Javier and Zayats, Anatoly V.},
	month = mar,
	year = {2017},
	pages = {133605},
}

@article{sanders_nanoscale_2019,
	title = {Nanoscale transfer of angular momentum mediated by the {Casimir} torque},
	volume = {2},
	copyright = {2019 The Author(s)},
	issn = {2399-3650},
	url = {https://www.nature.com/articles/s42005-019-0163-3},
	doi = {10.1038/s42005-019-0163-3},
	number = {1},
	urldate = {2026-08-29},
	journal = {Communications Physics},
	publisher = {Nature Publishing Group},
	author = {Sanders, Stephen and Kort-Kamp, Wilton J. M. and Dalvit, Diego A. R. and Manjavacas, Alejandro},
	month = jun,
	year = {2019},
	pages = {71},
}

@article{gao_thermal_2021,
	title = {Thermal equilibrium spin torque: {Near}-field radiative angular momentum transfer in magneto-optical media},
	volume = {103},
	shorttitle = {Thermal equilibrium spin torque},
	url = {https://link.aps.org/doi/10.1103/PhysRevB.103.125424},
	doi = {10.1103/PhysRevB.103.125424},
	number = {12},
	urldate = {2026-08-29},
	journal = {Physical Review B},
	publisher = {American Physical Society},
	author = {Gao, Xingyu and Khandekar, Chinmay and Jacob, Zubin and Li, Tongcang},
	month = mar,
	year = {2021},
	pages = {125424},
}

@article{raissi_physics-informed_2019,
	title = {Physics-informed neural networks: {A} deep learning framework for solving forward and inverse problems involving nonlinear partial differential equations},
	volume = {378},
	issn = {0021-9991},
	shorttitle = {Physics-informed neural networks},
	url = {https://www.sciencedirect.com/science/article/pii/S0021999118307125},
	doi = {10.1016/j.jcp.2018.10.045},
	urldate = {2026-08-29},
	journal = {Journal of Computational Physics},
	author = {Raissi, M. and Perdikaris, P. and Karniadakis, G. E.},
	month = feb,
	year = {2019},
	pages = {686--707},
}

\end{document}